\documentclass[fleqn,usenatbib,useAMS]{mnras}

\usepackage{graphicx}	
\usepackage{amsmath}	
\usepackage{amssymb}	
\usepackage{multicol}        
\usepackage{bm}		
\usepackage{pdflscape}	

\usepackage{multirow}
\usepackage{longtable}

\newcommand{\cfhqs}{CFHQS\,J142952+544717}
\newcommand{\srge}{SRGE\,J142952.1+544716}
\newcommand{\SRG}{{\it SRG}}

\newcommand{\eROSITA}{{\it eROSITA}}

\newcommand{\XMM}{{\it XMM-Newton}}

\newcommand{\gammax}{\Gamma_{\rm x}}
\newcommand{\alphax}{\alpha_{\rm x}}
\newcommand{\px}{p_{\rm x}}
\newcommand{\alphar}{\alpha_{\rm r}}
\newcommand{\pr}{p_{\rm r}}
\newcommand{\lc}{L_{\rm c}}
\newcommand{\ld}{L_{\rm d}}

\usepackage[T1]{fontenc}
\usepackage{ae,aecompl}

\usepackage{newtxtext,newtxmath}

\title[XMM-Newton observations of CFHQS J142952+544717]{XMM-Newton observations of the extremely X-ray luminous quasar CFHQS J142952+544717=SRGE J142952.1+544716 at redshift z=6.18}
\author[Medvedev et al.]{Medvedev P.$^1$
\thanks{Contact e-mail: \href{mailto:tomedvedev@iki.rssi.ru}{tomedvedev@iki.rssi.ru}},
Gilfanov M.$^{1,2}$,
Sazonov S.$^{1,3}$,
Schartel N.$^4$,
Sunyaev R.$^{1,2}$
\\
$^1$Space Research Institute (IKI), Russian Academy of Sciences, Profsoyuznaya ul. 84/32, Moscow, 117997 Russia \\
$^2$Max-Planck-Institut f\"{u}r Astrophysik (MPA), Karl-Schwarzschild-Str. 1, D-85741 Garching, Germany \\
$^3$Moscow Institute of Physics and Technology, Institutsky per. 9, 141700 Dolgoprudny, Russia \\
$^4$European Space Agency (ESA), European Space Astronomy Centre (ESAC), E-28691 Villanueva de la Cañada, Madrid, Spain
}

\date{Last updated ...; in original form ...}

\pubyear{2020}

\begin{document}
\label{firstpage}
\pagerange{\pageref{firstpage}--\pageref{lastpage}}
\maketitle

\begin{abstract}
We present results from a 20~ks \XMM\ DDT\ observation of the radio-loud quasar \cfhqs\ at $z=6.18$, whose extreme X-ray luminosity was recently revealed by the \SRG/\eROSITA\ telescope in the course of its first all-sky survey. The quasar has been confidently detected with a total of $\sim 1400$ net counts in the 0.2--10 keV energy band (1.4 to 72 keV in the object's rest frame). Its measured spectrum is unusually soft and can be described by an absorbed power-law model with a photon index of $\Gamma = 2.5\pm0.2$. There are no signs of a high-energy cutoff or reflected component, with an 90 per cent upper limit on the fluorescence iron K$\alpha$ equivalent width of $\approx 290$ eV and the corresponding upper limit on the iron K-edge absorption  depth of 0.6. We have detected,  at the $> 95$ per cent confidence level, an excess absorption above the Galactic value, corresponding to a column density $N_H=3\pm2 \times 10^{22}$ cm$^{-2}$ of material located at $z=6.18$. The intrinsic luminosity of \cfhqs\ in the 1.4 to 72 keV energy band is found to be $5.5_{-0.6}^{+0.8} \times 10^{46}$~erg~s$^{-1}$.  We did not detect statistically significant flux changes between two \SRG\ scans and the \XMM\ observation, spanning over $\sim 7.5$ months, implying that the quasar remained at this extremely high luminosity level for at least a month in its rest frame. We put forward the hypothesis that the extreme X-ray properties of \cfhqs\ are associated with inverse Compton scattering of cosmic microwave background photons (at $z=6.18$) in its relativistic jets.
\end{abstract}

\begin{keywords}
galaxies: active, galaxies: high-redshift, galaxies: nuclei, X-rays: general, individual: \cfhqs
\end{keywords} 



\section{Introduction}
\label{sec:intro}
\cfhqs\ is a radio-loud, optically luminous distant quasar at a redshift of 6.18. \cfhqs\ belongs to the bright end of the quasar luminosity function at $z\sim 6$ (e.g. \citealt{Willott_CFHQS}), but is not extremely bright, being $\sim 1.5$ magnitudes fainter than the most optically luminous quasars known at these redshifts.  With a radio-loudness parameter of $R\sim100$, the quasar has a steep radio spectrum with no significant variability detected in the radio (for detailed discussion, see \citealt{Medvedev2020}). According to these properties, relativistic beaming is unlikely to play a major role in the appearance of the source, and therefore \cfhqs\ is unlikely to be a blazar.

Recently, \cite{Medvedev2020} have discovered an X-ray signal from \cfhqs\ by cross-matching the catalogue of sources detected in the first \SRG/\eROSITA\ all-sky survey with the Pan-STARRS1 (PS1) distant quasar sample in the redshift range of $5.6 < z < 6.7$ \citep{Banados_distant_quasars}. With the measured luminosity of $2.6^{+1.7}_{-1.0}\times 10^{46}$~erg~s$^{-1}$ in the rest-frame 2--10~keV energy band, the object (called \srge\ in the \SRG/\eROSITA\ catalogue) turned out to be the most X-ray luminous quasar known at $z>6$\footnote{A second quasar with comparably high X-ray luminosity was recently discovered by \SRG/\eROSITA\ at a slightly smaller redshift of $z=5.47$ \citep{Khorunzhev2021}}. 
Because of the small number ($\sim 10$) of photons detected by \eROSITA, only a crude spectral analysis was done by \cite{Medvedev2020}, which provided loose constraints on the power-law photon index: $\Gamma =1.4 \pm 0.9$. 

Following the discovery of the extreme X-ray luminosity of this quasar, an \XMM\ \citep{xmm} Director’s Discretionary Time (DDT) observation of \cfhqs ~was scheduled in the summer 2020, aimed at obtaining a high-quality X-ray spectrum of the source and investigating its short- and long-term scale variability. The results of these \XMM\ observations as well as from the second scan of this source by \SRG/\eROSITA\ during its ongoing all-sky survey are reported below. 

In what follows, we adopt a flat $\Lambda$CDM cosmological model with $h = 0.70$ and $\Omega_{\Lambda} = 0.7$, and the quasar's redshift $z=6.183$ \citep{Wang_CO}.

\section{Data reduction and methods}
\label{sec:data}

\begin{figure}
\centering
\includegraphics[width=0.9\columnwidth]{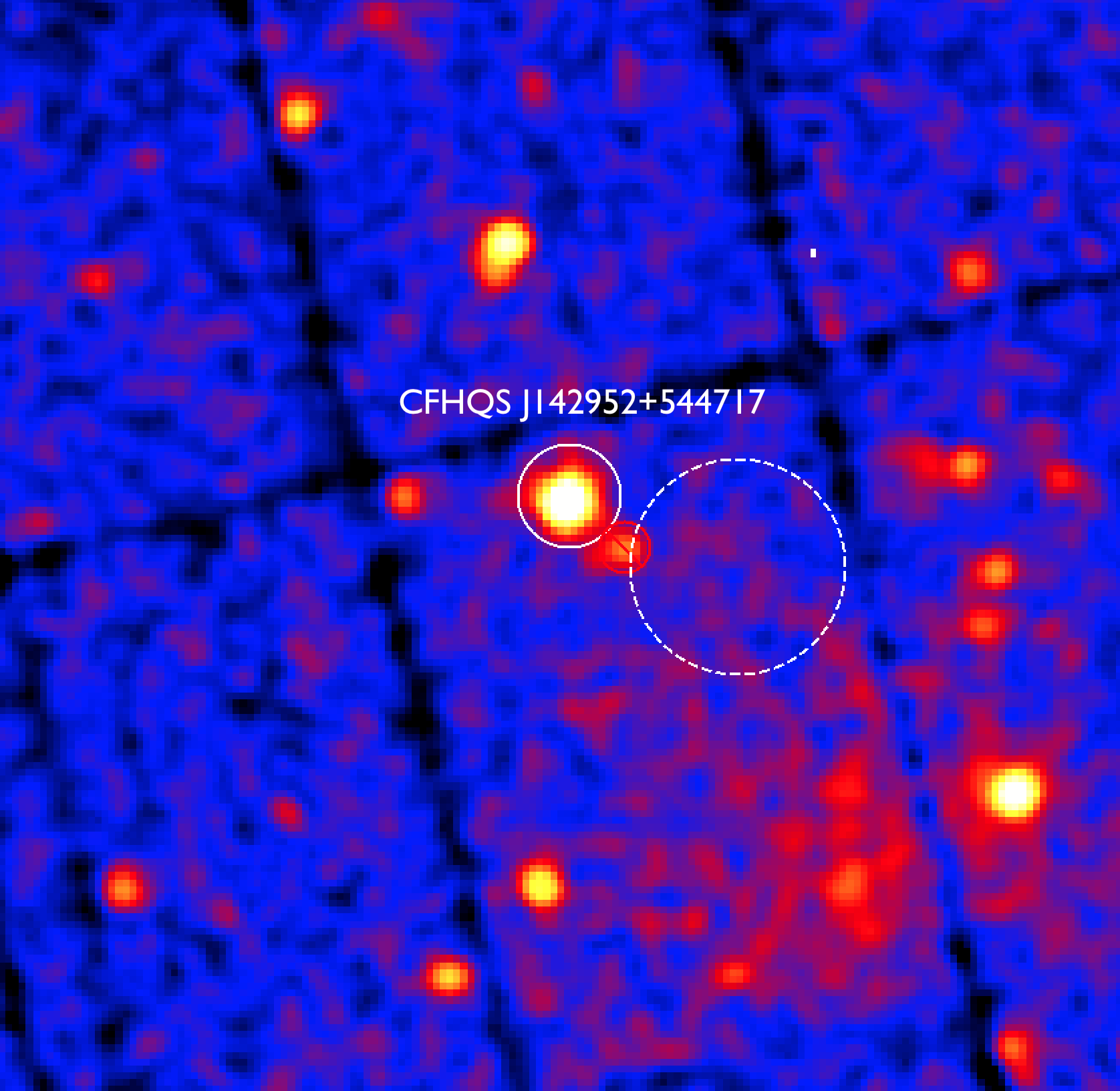}
\caption{Mosaic image of \cfhqs\ = \srge\ obtained by \XMM, including all three EPIC cameras, in the 0.2--12 keV band, smoothed with a Gaussian filter with $\sigma=1.5$. The white circles show extraction regions for the PN spectra: the solid circle of 30~arcsec radius is the source region and the dashed circle of $\approx60$ arcsec radius is the background region. The red circle (12 arcsec) masks a faint source and was excluded from the background extraction region.}
\label{fig:epic}
\end{figure}
\cfhqs\ was observed with \XMM\  on 2020 July 24 for 23 ks of Director’s Discretionary Time (revolution 3777, obsID 0871191201). The European Photon Imaging Camera (EPIC, \citealt{epic_pn, epic_mos}) was operated in full-frame mode with thin filters. The event lists for the EPIC cameras were obtained from the processed pipeline products (PPS) retrieved from the \XMM\ Science Archive. As a first step, we selected all events that were not affected by proton flares. The total cleaned exposure times are 19.7 and 21 ks for the PN and MOS cameras, respectively.

We extracted the source spectra using a circle region with a radius of 30 arcsec centred on the best-fitting source position from the maximum-likelihood source list (ML-list) for PSF fitting provided by PPS. For the background extraction, we used a 60 arcsec circle avoiding CCD gaps for PN (see Fig.~\ref{fig:epic}) and 3 circles (of 100, 90, and 70 arcsec) in a source free area for the MOS cameras. For each spectrum we generated the response matrix and the ancillary file using the Science Analysis System (SAS v.18.0.0) tasks \texttt{RMFGEN} and \texttt{ARFGEN}. Event patterns 0--12 were included for the MOS cameras, while for the PN camera we used patterns 0--4. The net source count rates obtained are $1.0\pm 0.1 \times 10^{-2}$, $1.0\pm 0.1 \times  10^{-2}$ and $4.6\pm 0.2 \times 10^{-2}$ cts s$^{-1}$ for MOS-1, MOS-2 and PN, respectively. We did not consider the RGS data, since the source is too faint ($\ll 0.1$ cts s$^{-1}$) for a meaningful RGS spectral analysis.

The spectra were rebinned to ensure a minimum of 5 counts per energy bin by means of the standard tool \texttt{GRPPHA}. Then we use the C-statistic \citep{cstat} modified to account for the Poisson background subtraction (the `W-statistic' in \texttt{XSPEC}, version 12.11.0, \citealt{Arnaud1996}) to analyse the data.  Once the best-fitting model is obtained, we run a Monte Carlo Markov Chain (MCMC) within XSPEC, using the Goodman-Weare algorithm \citep{emcee}  with $10^5$ steps to  explore the parameter space. All subsequent errors are calculated using MCMC chains and quoted at the 90 per cent confidence level. We also examine the goodness-of-fit for individual models by using the `goodness' command in \texttt{XSPEC} with Anderson-Darling (AD) test statistic (see, e.g., Appendix B in the \texttt{XSPEC} manual for definition).  This test reports the fraction of simulations having an AD statistic smaller than that for the observed spectrum. In what follows, we run this command with  `nosim' and `fit' options, which means that all simulations are drawn from the best-fitting model; then each simulated data set is fitted to the model and new test statistic is calculated (also referred to as the parametric bootstrap method, see, e.g. \citealt{feigelson2012}). For the `goodness' test, we run 100,000 realisations for each model. As is well known, the goodness-of-fit test result should not be interpreted as the probability of the model being (in)correct, rather it helps spot those models whose best-fitting statistic values are too high to be a result of a statistical fluctuation.

\section{Results}
\label{sec:res}
\cfhqs\ is clearly detected in the \XMM\ EPIC images (see Fig.~\ref{fig:epic}), with $\approx 1400$ net counts from the source in the 0.2--12 keV energy range. The spectrum is relatively soft, but the source is nevertheless confidently detected in the 4.5--7.5 keV band. 
 In this energy range, we detected a total of 72 counts in the source extraction region, with the expected background contribution of 52.9 counts (the Poisson probability is $\ln(p) = -5.3$, corresponding to $\approx 2.8\sigma$ for Gaussian distribution).
No detection was obtained in the 7.5--12 keV band at 90 per cent confidence; given a total of 69 registered counts with the expected background contribution of 61.2 counts ($\ln(p) = -1.9$), the corresponding upper limit flux is $5.5  \times 10^{-14}$ erg cm$^{-2}$ s$^{-1}$.

\begin{figure}
\centering
\includegraphics[width=1.\columnwidth]{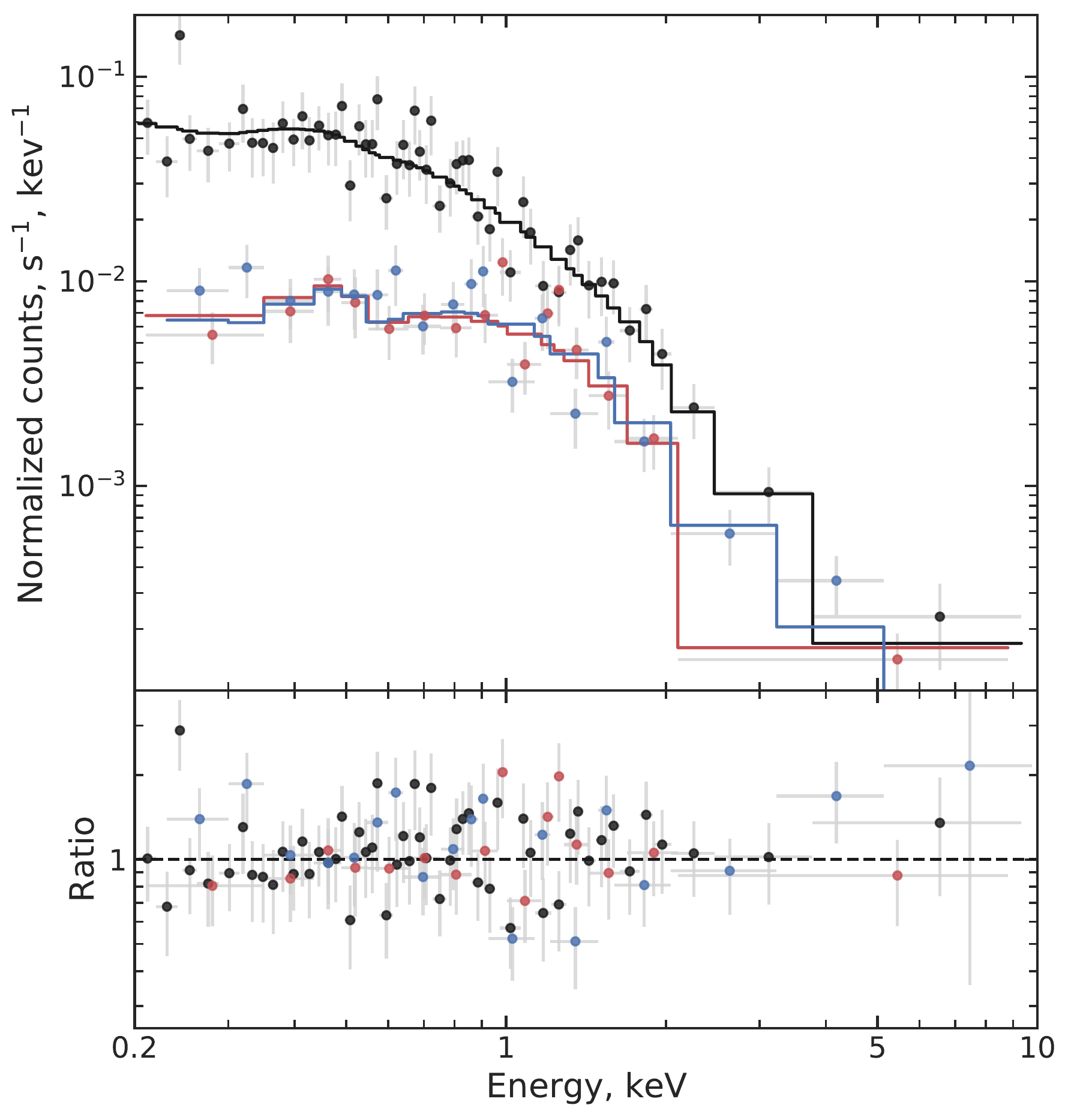}
\caption{EPIC-PN (black) and EPIC-MOS (red and blue) spectra of \cfhqs. The spectral bins are combined to have a significant detection of at least 3$\sigma$ (for plotting purposes only). The solid lines show an absorbed power-law model providing the best simultaneous fit to the data in the 0.2--10 keV range (see model 1 in Table~\ref{tab:bf}). The bottom panel shows the ratio of the data to the folded model. The shown observed energy range corresponds to the 1.4--72 keV range in the quasar's rest frame. }
\label{fig:spec}
\end{figure}

Since the \XMM\ astrometric accuracy from the attitude solutions is subject to systematic uncertainties, typically $\sim 1$ arcsec \citep[e.g., ][]{XMM_PPS}, the cross-match of the EPIC sources from the ML-list with sources from DSS-DR9 is performed during the pipeline data processing. In the case of the \cfhqs\  observation, we found offsets in the RA, DEC position of 0.97 and $-1.47$ arcsec, respectively. The corrected X-ray source position is $\alpha=14^{h}29^{m}52.188^{s}$, $\delta=+54^{d}47^{m}17.86^{s}$ with the statistical error $\mathrm{RADEC\_ERR}=0.25$ arcsec and the 
error arising from the field rectification $\mathrm{SYSERRRCC}=0.39$ arcsec, which combined can be translated to a $1\sigma$ positional uncertainty of 0.5 arcsec. The obtained position is 0.5~arcsec offset from the optical position of the quasar \cfhqs{} ($\alpha=14^{h}29^{m}52.146^{s}$, $\delta=+54^{d}47^{m}17.55^{s}$) in CFHTLS \citep{Hudelot_CFHTLS}, while the nearby serendipitous source CFHT\,1323\_175499 discussed in \cite{Medvedev2020} is 2.1 arcsec away from the X-ray source. We thus confirm the conclusion drawn in \cite{Medvedev2020} that the X-ray source \srge\ is indeed associated with the quasar \cfhqs.

On the \XMM\ image (Fig. \ref{fig:epic}), there is a weak X-ray source located at $\alpha=14^{h}29^{m}48.246^{s}$, $\delta=+54^{d}46^{m}50.28^{s}$,  $\approx 40\arcsec$ to the south-west of \cfhqs. This source has a counterpart in SDSS with $z=21.09\pm 0.44\ \mathrm{mag}$, $i=21.95\pm 0.25\ \mathrm{mag}$  (shown by the red circle). Its optical colours suggest that it is a relatively nearby object and is therefore unrelated to \cfhqs.

\subsection{X-ray spectrum}
\label{sec:spec}

The $\sim 1400$ net counts detected from the source enable its detailed spectral analysis. We fitted the individual EPIC camera spectra simultaneously in the 0.2--10 keV energy range, adding cross-normalization factors between the cameras and allowing them to vary. The results of the spectral fitting are summarized in Table~\ref{tab:bf}.

We started by fitting the data with an absorbed power-law model. We first fixed the neutral hydrogen column density at the Galactic value in the direction of \cfhqs, $N_{\rm H}=1.15\times10^{20}$~cm$^{-2}$ \citep{HI4PI_collab} and obtained an inadequate fit with the cstat value of 279 for 260 d.o.f and the goodness of the fit value of 99  per cent. We therefore thawed the absorption and repeated the fit again. The folded EPIC spectra along with the best-fitting model are shown in Fig.~\ref{fig:spec}, the best-fitting parameters are summarized in  Table~\ref{tab:bf}. We obtained a fairly soft power-law spectrum with a photon index of $\Gamma = 2.5 \pm 0.2$ and an absorption column density of $N_H = 4 \pm 2 \times10^{20}$~cm$^{-2}$. The Galactic absorption column is thus outside of the error interval at more than 90 per cent confidence. To examine the quality of the fit, we ran a `goodness' simulation (see Section~\ref{sec:data}). It showed that 74 per cent of the realisations have the value of AD statistic smaller than that obtained for the real data, suggesting that the model provides an acceptable fit to the observed spectrum.

Using the best-fitting model, we find that the X-ray luminosity of \cfhqs\ is $L_X = 3.0_{-0.4}^{+0.6} \times 10^{46}$~erg~s$^{-1}$ in the rest-frame 2--10~keV energy band, in good agreement with the previous measurement by \SRG/\eROSITA\ \citep{Medvedev2020}. The luminosity in the entire observed energy range (1.4--72 keV in the rest frame) is $5.5_{-0.6}^{+0.8} \times 10^{46}$~erg~s$^{-1}$. 

As a next step, we searched for evidence of a high-energy roll-over in the spectrum. To this end, we replaced the power-law component by the \textsc{zcutoffpl} model  in \texttt{XSPEC} with the redshift fixed at 6.183 and the absorption column density fixed at the Galactic value. We found no significant evidence for continuum curvature at the high-energy end of the spectrum with a 90 per cent lower limit on the $e$-folding energy of the exponential cut-off of $\beta=29$ keV ($\approx 4$ keV in the observer's frame). As no extra absorption was included in the model, the fitting resulted in a flatter photon index  of $\Gamma=2.2_{-0.2}^{+0.1}$. The model gives a slightly worse fit statistics, by $\Delta C$-statistic$ = 6$ for the same value of $d.o.f = 259$, confirming a preference for some additional absorption by the data. Running a `goodness' test, we found that 97 per cent of the realisations have a lower AD statistic.
 
\begin{figure}
\centering
\includegraphics[width=1.\columnwidth]{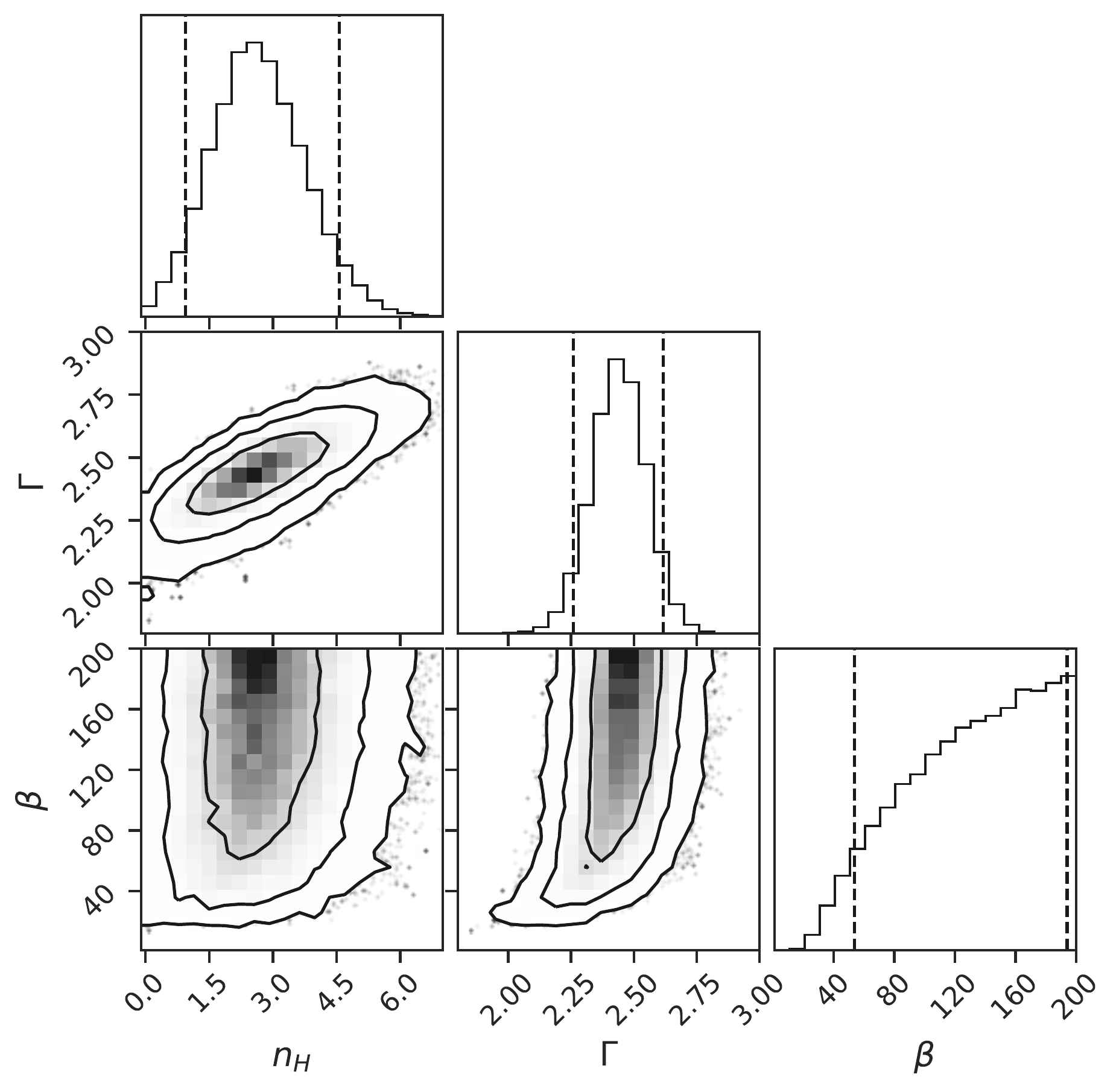}
\caption{One- and two-dimensional projections of the posterior probability distributions over 3 parameters of model 3 (see Table~\ref{tab:bf}): intrinsic column density ($10^{22}$ cm$^{-2}$), power-law photon index and high-energy cut-off (rest frame, keV). Contours represent the 68, 95 and 99.7 per cent probability regions, sample points outside of these regions being plotted as gray dots. The vertical dashed lines show the 0.05 and 0.95 quantiles (90 per cent confidence).}
\label{fig:nH}
\end{figure}

We further added an intrinsic absorption by material located at $z=6.183$ to the absorbed cutoff power-law model. The posterior probability distributions and confidence contours for the photon index $\Gamma$, intrinsic neutral hydrogen column $N_H$ and cutoff energy $\beta$ are shown in Fig.~\ref{fig:nH}. Comparing the Akaike Information criterion \citep{AIC} for this model with that for model 1 (see Table~\ref{tab:bf}), we can again confirm that the high energy cut-off is not required by the data, the model with the cutoff yielding almost the same best-fitting statistics but with an increased number of free parameters. In contrast, the intrinsic absorption in excess of the Galactic value is detectable with more than 95 per cent confidence, with the best-fitting value of $N_{\rm H}=3\pm 2 \times10^{22}$~cm$^{-2}$.

\begin{table*}
\centering
\caption{Results of the spectral analysis for the \XMM\ observation of \cfhqs. In all models, the redshift is fixed at 6.183 if applicable. The abundance set by \protect\cite{wilms2000} is used in all absorption models. The quoted broad-band (0.2--10 keV) flux is the flux for the best-fitting model to PN-data corrected for the Galactic and intrinsic absorption. $N_{M1}$ and $N_{M2}$ stand for the MOS1-to-PN and MOS2-to-PN cross-calibration factors, respectively. The cstat values correspond to the negative of twice the logarithm of the likelihood of the model (by definition). The `goodness' values, below the AD statistic values, denote the percentage of realisations of the model that have a lower AD statistic than the data (obtained by the `goodness' command in \texttt{XSPEC}). Thus, larger `goodness' values indicate a smaller probability of obtaining such a poor fit by chance. The Akaike information criterion (AIC) is calculated as $AIC = 2k + cstat$, where $k$ is the numbed of free parameters in the fit, the preferred model for the data being the one with the minimum AIC.}
\label{tab:bf}
\begin{tabular}{c  c  c  c  c  c}
\hline
\multirow{2}{*}[0.0cm]{\bf Xspec Model}  &	\multirow{2}{*}[0.0cm]{\bf Parameters} & \multirow{2}{*}[0cm]{\bf Parameter Values} &	\multirow{2}{*}[0cm]{{\bf cstat}/{\bf d.o.f}} & {\bf log(AD)} & \multirow{2}{*}[0cm]{\bf AIC} \\
 & & & & {\bf (goodness)} \\
\hline
\multirow{4}{1in}[-0.3cm]{1. \textsc{const*tbabs*pow}} & $N_H$ &	$4\pm2 \times 10^{20}$ cm$^{-2}$  & \multirow{4}{*}[-0.3cm]{272/259} & \multirow{4}{0.5in}[-0.3cm]{\centering $-7.48$ \\$(73.9 \%)$} & \multirow{4}{*}[-0.3cm]{281.6} \\
    [0.1cm] &  $\Gamma$  & $2.5\pm0.2$  &	& \\
    [0.1cm] &  $N_{M1}$  & $1.1_{-0.1}^{+0.2}$ & & \\
    [0.1cm] &  $N_{M2}$  & $1.2\pm0.2$ & & \\
    [0.1cm] &  $F_{0.2-10}$  & $1.3_{-0.1}^{+0.2} \times 10^{-13}$ erg cm$^{-2}$ s$^{-1}$ & & \\[0.1cm]
\hline

\multirow{6}{1in}[-0.3cm]{2. \textsc{const*tbabs*\\zcutoffpl}} &	$N_H^{*}$ & $1.15 \times 10^{20}$ cm$^{-2}$ & \multirow{6}{*}[-0.3cm]{278/259} & \multirow{6}{0.5in}[-0.3cm]{\centering $-6.68$ \\$(97.4\%)$} & \multirow{6}{*}[-0.3cm]{288.1} \\
    [0.1cm] & $\Gamma$ & $2.2_{-0.2}^{+0.1}$ & & \\
    [0.1cm] & $\beta$ & $>29$ keV & & \\
    [0.1cm] &  $N_{M1}$  & $1.1_{-0.1}^{+0.2}$ & & \\
    [0.1cm] &  $N_{M2}$  & $1.2\pm0.2$ & & \\
    [0.1cm] & $F_{0.2-10}$ & $1.08_{-0.10}^{+0.09} \times 10^{-13}$ erg cm$^{-2}$ s$^{-1}$ & & \\[0.1cm]
    \hline

\multirow{7}{1in}[-0.3cm]{3. \textsc{const*tbabs*\\zphabs*zcutoffpl}} & $N_H^{*}$ & $1.15 \times 10^{20}$ cm$^{-2}$ &	\multirow{7}{*}[-0.3cm]{$271/258$}  & \multirow{7}{0.5in}[-0.3cm]{\centering $-7.23$ \\$(86.9\%)$} & \multirow{7}{*}[-0.3cm]{283.5}\\
    [0.1cm] & $N_H^{z}$ & $3 \pm 2 \times 10^{22}$ cm$^{-2}$ &	& \\
    [0.1cm] & $\Gamma$ & $2.5\pm0.2$ &	& \\
    [0.1cm] & $\beta$ & $>54$ keV & & \\
    [0.1cm] &  $N_{M1}$  & $1.1_{-0.1}^{+0.2}$ & & \\
    [0.1cm] &  $N_{M2}$  & $1.1\pm0.2$ & & \\
    [0.1cm] & $F_{0.2-10}$ & $1.2_{-0.1}^{+0.2} \times 10^{-13}$ erg cm$^{-2}$ s$^{-1}$ & & \\[0.1cm]
    \hline
    
\multirow{8}{1in}[-0.3cm]{4. \textsc{const*tbabs*\\zphabs*\\(zedge*pow + zgauss)}} & $N_H^{*}$ & $1.15 \times 10^{20}$ cm$^{-2}$ & \multirow{8}{*}[-0.3cm]{$267/257$} & \multirow{8}{0.5in}[-0.3cm]{\centering $-7.76$ \\$(55.3\%)$} & \multirow{8}{*}[-0.3cm]{281.5}\\
    [0.1cm] & $N_H^{z}$ & $3 \pm 2 \times 10^{22}$ cm$^{-2}$ & & \\
    [0.1cm] & $\Gamma$	& $2.5\pm0.2$ & & \\
    [0.1cm] & $\tau_{\mathrm{edge}}$ & $0.3_{-0.2}^{+0.3}$ & & \\
    [0.1cm] & $EW_{\mathrm{Fe\,K}\alpha}$ &	$<40$ eV & & \\
    [0.1cm] & $\sigma_{\mathrm{Fe\,K}\alpha}^{*}$ &	$10$ eV & & \\
    [0.1cm] &  $N_{M1}$  & $1.1_{-0.1}^{+0.2}$ & & \\
    [0.1cm] &  $N_{M2}$  & $1.1\pm0.2$ & & \\
    [0.1cm] & $F_{0.2-10}$ & $1.3_{-0.1}^{+0.2} \times 10^{-13}$ erg cm$^{-2}$ s$^{-1}$ & & \\[0.1cm]
    \hline
\end{tabular}\\
$^{*}$ Parameter is fixed during the fit.
\end{table*}

\begin{figure}
\centering
\includegraphics[width=1.\columnwidth]{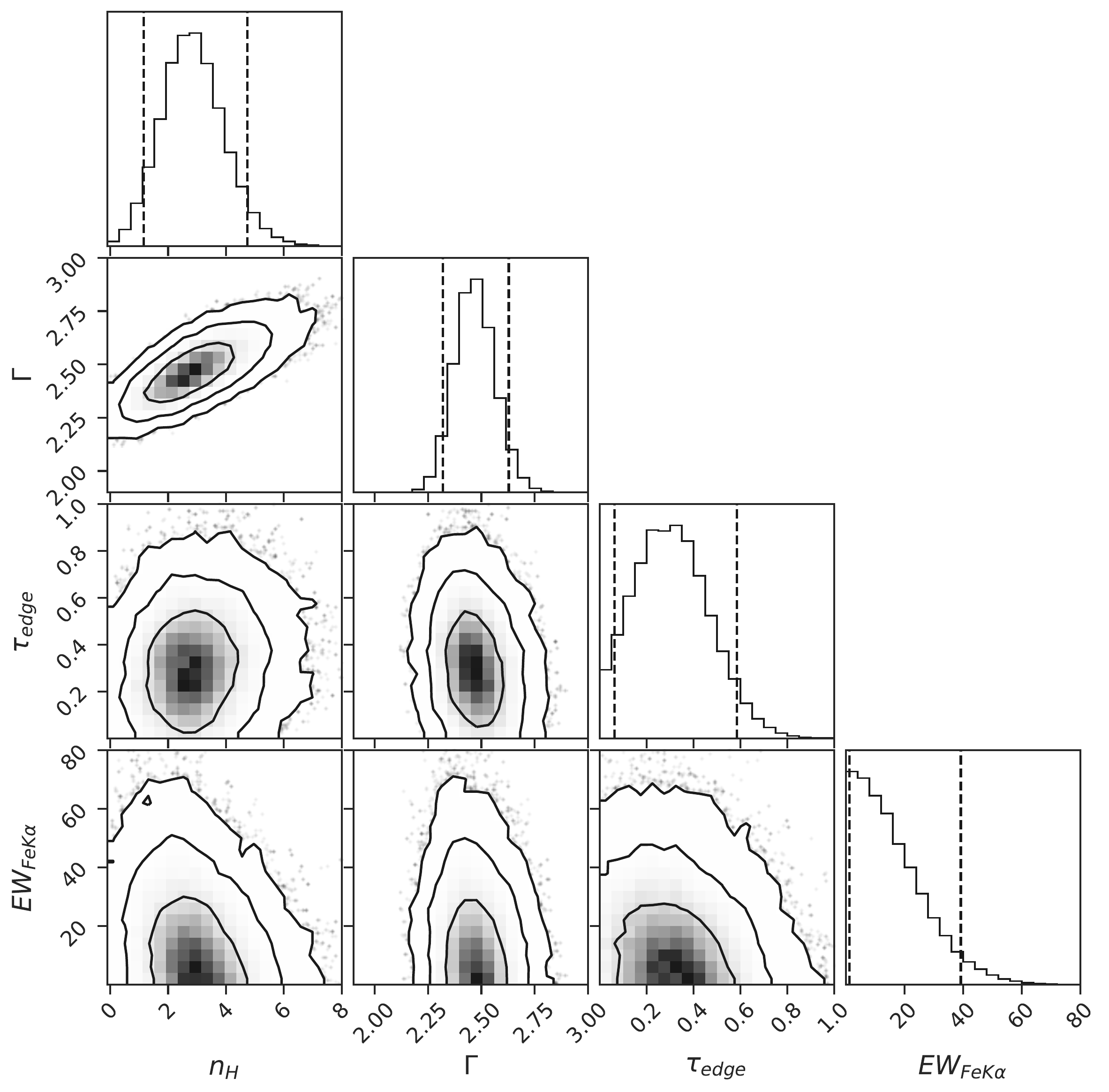}
\caption{Corner plot over 4 parameters of model 4 (see Table~\ref{tab:bf}): intrinsic column density ($10^{22}$ cm$^{-2}$), power-law photon index, absorption depth at rest-frame energy 7.1 keV (or 0.99 keV in the observer's frame) and equivalent width of Fe\,K$\alpha$ line at rest-frame energy 6.4 keV (or 0.9 keV in the observer's frame, eV). Contours represent the 68, 95 and 99.7 per cent probability regions, sample points outside of these regions being plotted as gray dots. The vertical dashed lines show the 0.05 and 0.95 quantiles (90
per cent confidence).
}
\label{fig:edge}
\end{figure}

Finally, we searched for signs of a reflection component, namely, the iron K$\alpha$ fluorescent emission line and the corresponding absorption edge. To  this end, we added to the absorbed power-law model a narrow line ($\sigma=10$ eV) centred at 6.4 keV and an absorption edge at 7.1 keV in the quasar's rest frame. We found no evidence for an iron line with a 90 per cent upper limit on its equivalent width of 40 eV  in the observer's frame, which translates to $\approx 290$ eV in the quasar's rest frame. For the depth of the absorption edge, we obtained a non-zero  best-fitting value of $\tau_{\mathrm{edge}} = 0.3^{+0.3}_{-0.2}$, which, however, is consistent with zero at the $\sim 95$ per cent confidence level, see Fig.~\ref{fig:edge}. We conclude that no spectral features associated with iron absorption and fluorescence have been detected in the data.


\subsection{Light curve}
\label{sec:lc}

\begin{figure}
\centering
\includegraphics[width=1\columnwidth]{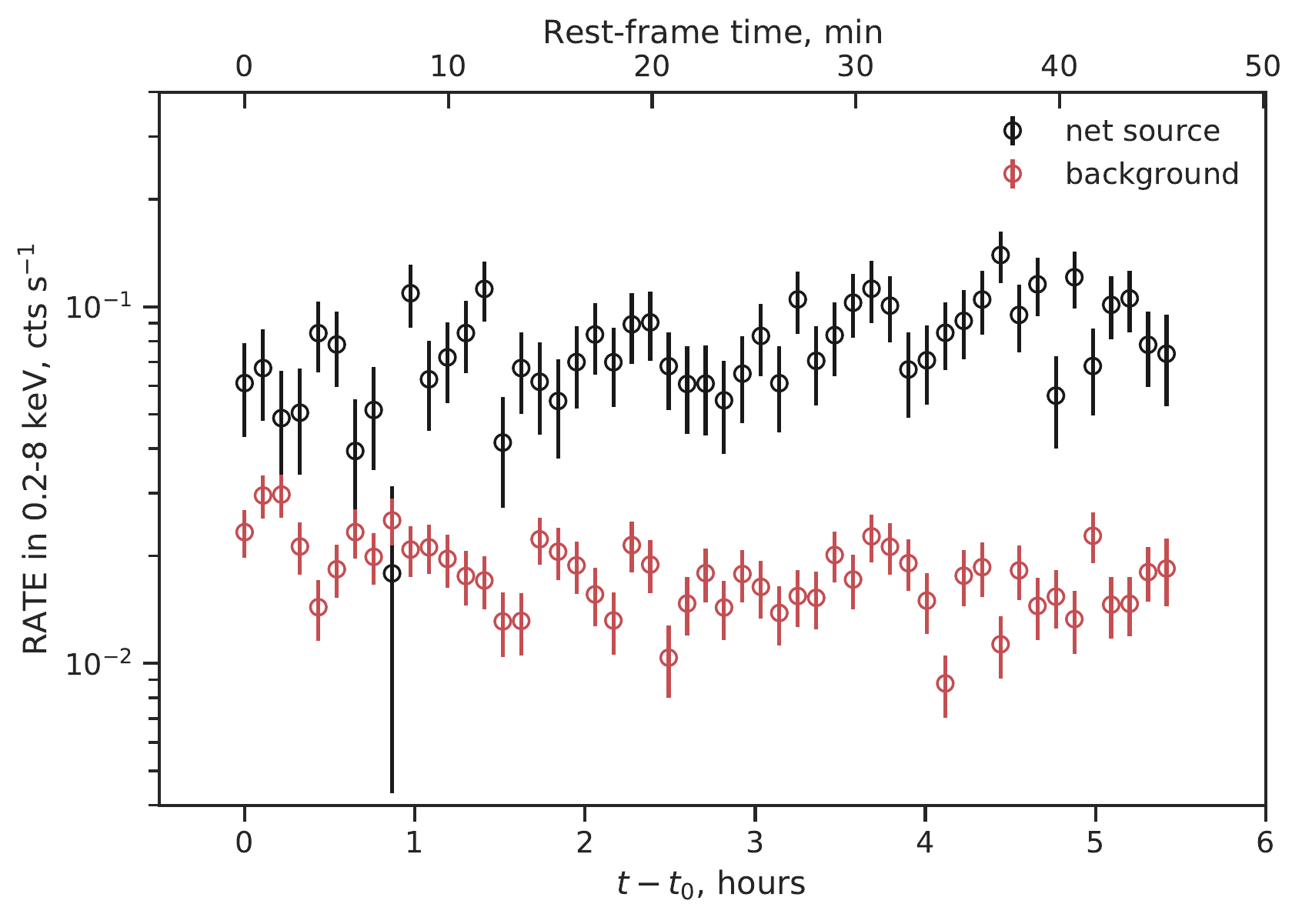}
\caption{Combined EPIC-PN and -MOS light curves of \cfhqs\ in the 0.2--8 keV energy ranges. 
The horizontal axis shows time since the start of the observation on 2020-07-24 12:27 UTC (in hours), the time bin is 390 s. The top axis shows time in the quasar’s rest frame (in minutes). The black circles show background subtracted source count rates, while the red circles show background rates. }
\label{fig:lc_rate}
\end{figure}

Figure~\ref{fig:lc_rate} shows the EPIC cameras combined, background-subtracted light curve of \cfhqs\ in the 0.2--8 keV energy range. We did not detect any statistically significant short-term variability, with the Fourier power spectrum being consistent with zero. Overall, the \XMM\ light curve may suggest some increasing trend, however, its statistical significance is low. Indeed, dividing the data into two halves and using model 1 from Table~\ref{tab:bf}, we measure the 0.2--10 keV band flux of $F_{0.2-10} = 1.17_{-0.15}^{+0.19} \times 10^{-13}$ and $F_{0.2-10} = 1.49_{-0.18}^{+0.24} \times 10^{-13}$ erg cm$^{-2}$ s$^{-1}$ for the first and second half, respectively .

These two values are consistent with each other within 90 per cent confidence. Thus, although the intra-day  variability of \cfhqs\  is an interesting possibility, which might dramatically impact the theoretical models for this source, it cannot be established with the due statistical significance at this stage.

We also checked if \cfhqs\ is variable on timescales of months by comparing two \eROSITA\ measurements conducted during the first and the second \SRG\ all-sky surveys with the \XMM\ measurement. The quasar position was scanned by \SRG\ on 2019 December 10--11 during eRASS\,I and on 2020 June 6--9 during eRASS\,II. Using the best-fitting model 1 from  Table~\ref{tab:bf}, we find the absorption uncorrected fluxes $F^{\mathbf{eI}}_{0.2-6} = 1.1_{-0.5}^{+0.6} \times 10^{-13}$ ergs cm$^{-2}$ s$^{-1}$ and $F^{\mathbf{eII}}_{0.2-6} = 1.3 \pm 0.6 \times 10^{-13}$ ergs cm$^{-2}$ s$^{-1}$ in the 0.2--6 keV energy range for eRASS\,I and II, respectively. We note that these flux values are based on pre-flight absolute flux calibrations of the \eROSITA\ telescope.  The given errors are statistical only (90 per cent confidence).  The corresponding flux during \XMM\ observation is $F_{0.2-6} = 8.4_{-0.7}^{+0.6}  \times 10^{-14}$ ergs cm$^{-2}$ s$^{-1}$ and is thus consistent, within statistical and systematic uncertainties, with the \eROSITA\ measurements taken $\sim 7$ and $\sim 1.5$ months earlier. 
 
We conclude that no statistically significant variability of the source has been detected.  

\section{Discussion}
\label{sec:summary}

The \XMM\ observations have confirmed beyond any doubt the association of the X-ray source SRGE\,J142952.1+544716 discovered by \SRG/\eROSITA\ with the quasar \cfhqs\ at $z=6.18$. The flux measurements taken with eROSITA and \XMM\ at three epochs between Dec. 2019 and July 2020 indicate that the X-ray luminosity of \cfhqs\ remained at an extremely high and nearly constant level for at least a month in the quasar's rest frame. 

\begin{figure}
\centering
\includegraphics[width=1.\columnwidth]{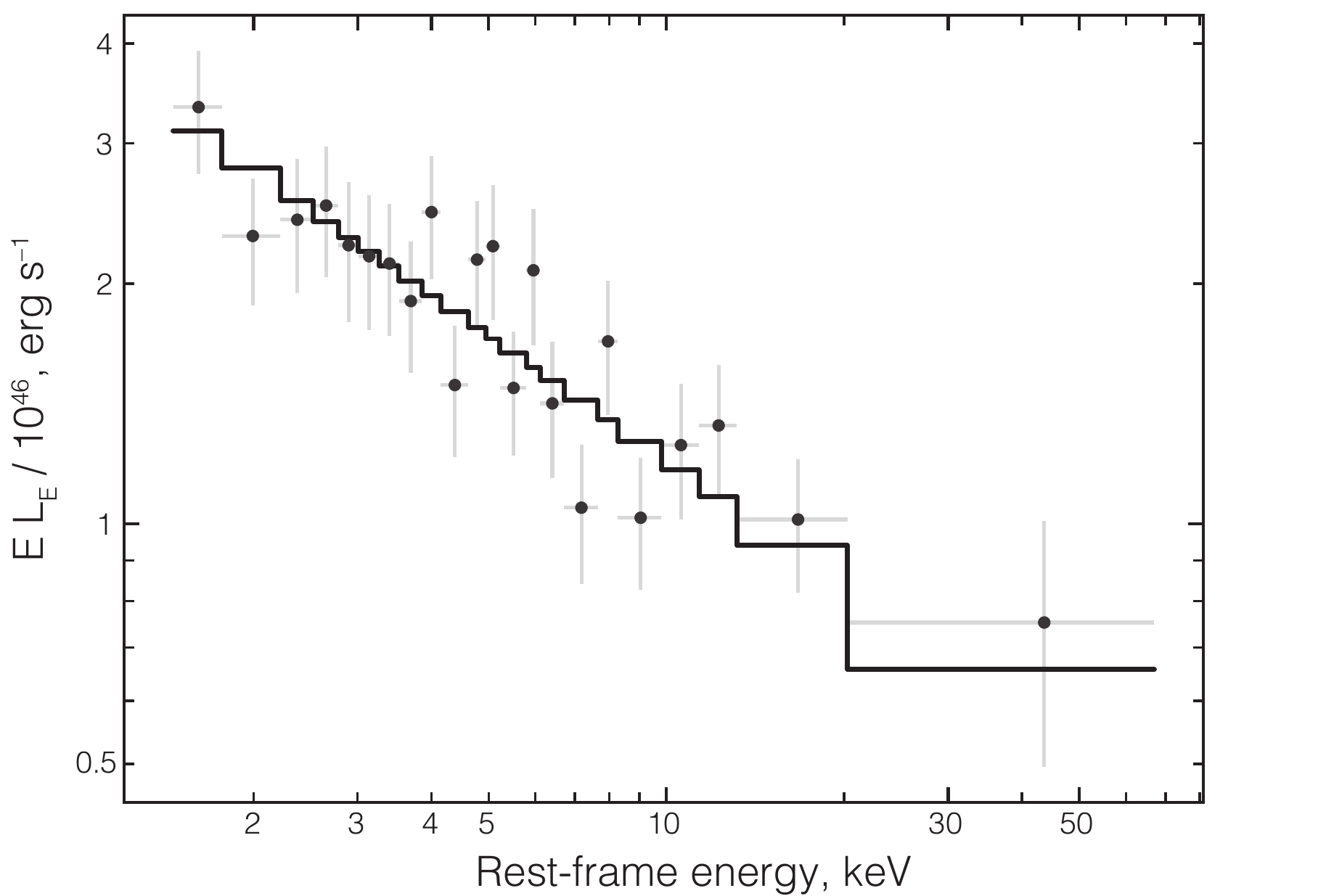}
\caption{\XMM\ EPIC-PN unfolded spectrum of \cfhqs\ in the quasar's rest frame. 
The spectrum was unfolded and unabsorbed using the best-fitting model 1 in Table~\ref{tab:bf}, which is shown by the black solid line. The spectral bins are combined to have at least 5$\sigma$ significance.
}
\label{fig:rest_spec}
\end{figure}

The X-ray spectrum measured with \XMM\ in the rest-frame 1.4--72~keV energy band has a simple power-law shape (slightly modified by absorption at low energies) without any significant features (Fig.~\ref{fig:rest_spec}). 
The steep slope of $\gammax=2.5\pm0.2$ is unusual for both Seyfert galaxies (e.g. \citealt{Malizia2014,Ricci2017}) and quasars (e.g. \citealt{Shemmer2008}), including those at $z\sim 6$ \citep{Nanni2017}, with the vast majority of the well-studied active galactic nuclei (AGN) having X-ray continua characterised by $\Gamma\sim 1.5$--2.2, usually interpreted in terms of Comptonization of accretion disc emission in a hot corona. 

The overall spectral energy distribution (SED) of \cfhqs, discussed in detail in \citet{Medvedev2020}, together with the refined X-ray spectrum reported here suggest that the SED's maximum is located between $\sim 10$~eV and $\sim 1$~keV (rest frame), i.e. in the UV-soft X-ray range. Within the accretion disc--corona scenario, this would imply that the luminosity of the Comptonization component is comparable to that of the thermal emission from the disc. Specifically, the corona's luminosity, $\lc$, is then at least $\approx 5\times 10^{46}$~erg~s$^{-1}$ (as directly measured by \XMM\ between rest-frame 1.4 and 72~keV) and more likely as high as $\sim 10^{47}$~erg~s$^{-1}$ (if we extrapolate the $\gammax\sim 2.5$ power-law spectrum down to 100~eV and up to $\sim 100$~keV); whereas the disc's luminosity, $\ld$, can be estimated by integrating the optical--UV part of the SED (see \citealt{Medvedev2020}) to be $\sim 10^{47}$~erg~s$^{-1}$. The inferred ratio $\ld/\lc\sim 1$ is within the range of typical values for Seyfert galaxies in the local Universe, $\ld/\lc\sim 1$--3, but lower than the corresponding range, $\ld/\lc\sim 3$--6, for typical, more luminous quasars in the Universe whose combined emission constitutes the bulk of the cosmic X-ray background \citep{Sazonov2004,Sazonov2012}. Moreover, many authors (e.g. \citealt{lusso10}) have reported a trend of the SED's optical-to-X-ray (between 2500\AA\ and 2~keV) effective slope increasing with increasing AGN luminosity, which would suggest that the $\ld/\lc$ ratio should be particularly high for such a luminous quasar as \cfhqs\ in contrast  to what is observed.


We thus conclude \citep[cf.][]{Medvedev2020} that \cfhqs\ deviates in its SED characteristics from the trends typically observed for quasars. The unusually steep X-ray spectrum and high X-ray/optical luminosity ratio indicate that the accretion disc's corona is probably not responsible for the bulk of the X-ray emission observed from \cfhqs. 

\subsection{X-ray jets}
\label{sec:jets}

We suggest that the extreme X-ray luminosity of \cfhqs\ (both the absolute value and the X-ray/optical ratio) is linked with its relativistic jets, the existence of which is evident from the fact that \cfhqs\ is a radio-loud quasar. Specifically, the observed X-ray emission might be produced by inverse Compton scattering of cosmic microwave background (CMB) photons off relativistic electrons in the jets. 

As discussed in \citet{Medvedev2020}, according to its multi-wavelength properties (in particular, the fairly steep spectrum and low brightness temperature of the radio emission as well as weak variability in all the observed energy ranges from radio to X-rays), \cfhqs\ is unlikely to be a blazar, i.e. its jet is not pointing at us. In order to make some crude estimates below, we assume that the jets have a bulk Lorentz factor $\Gamma=5$ and are oriented at a moderate angle $\cos\alpha\approx 1-1/\Gamma$ (i.e. $\alpha \approx 40^\circ$ for $\Gamma=5$) with respect to our line of sight so that the bulk Doppler factor $\delta=1/\Gamma(1-\beta\cos\alpha)\sim 1$ and can be omitted from the consideration.


The typical Lorentz factors, $\gamma$, of the electrons that up-scatter CMB photons to the X-ray range can be estimated from the standard relation $E(1+z)\sim (4/3)\Gamma\gamma^2\langle h\nu\rangle$,  which takes into account the fact that the incident CMB photons are typically blueshifted by a factor $\sim\Gamma$ due to the relativistic motion of the jet. Here, $\langle h\nu\rangle \approx 2.8 kT_{\rm CMB}(1+z)/h$ is a typical energy of the CMB photons, which can be taken to be equal to the peak energy of the Planck spectrum, $T_{\rm CMB}=2.725$~K is the CMB temperature at $z=0$, $E$ is the observed energy of the up-scattered photons, and $z=6.18$ is the object's redshift. Given that the X-ray spectrum measured with \XMM\ is a power law with $\alphax=1.5\pm 0.2$ ($dF_E/dE\propto E^{-\alphax}$) between $E_{\rm min}=0.2$~keV and $E_{\rm max}=10$~keV, we infer that the energy distribution of the underlying population of relativistic electrons extends down to at least $\gamma_{\rm min}\sim 500/\Gamma\sim 100$ and up to at least $\gamma_{\rm max}\sim 3000/\Gamma\sim 600$ and its energy distribution ($dN_{\rm e}/d\gamma\propto\gamma^{-\px}$) has a slope of $\px=2\alphax+1=4\pm 0.4$ \footnote{In view of other significant uncertainties in the presented scenario, we ignore here the possible impact of the exact shape of the relativistic Compton scattering kernel on these crude estimates.}. The cooling time of such electrons due to the inverse Compton scattering of the CMB photons at $z=6.18$ can be found as:
\begin{equation}
    t_{\rm cool}=\frac{3m_{\rm e}c^2}{16\sigma_{\rm T}\sigma T_{\rm CMB}^4(1+z)^4\Gamma^2\gamma}\approx 6\times 10^6\left(\frac{\Gamma}{5}\right)^{-2}\gamma^{-1}~{\rm yr},
    \label{eq:tcool}
\end{equation}
where $\sigma_{\rm T}$ is the Thomson cross-section, $\sigma$ is the Stefan--Boltzmann constant, and we have taken into account that the CMB energy density is enhanced by a factor of $\Gamma^2$ in the comoving frame of the jet (e.g. \citealt{ghisellini2009}). We find that $t_{\rm cool} (\gamma_{\rm min})\sim 3\times 10^5/\Gamma\sim 6\times 10^4$~yr and $t_{\rm cool} (\gamma_{\rm max})\sim 5\times 10^4/\Gamma\sim 10^4$~yr.

The X-ray jet will thus have a size of $\lesssim 20$~kpc (the distance travelled at a speed close to that of light over the cooling time of the lowest-energy X-ray emitting electrons). For viewing angles $\alpha\approx 40^\circ$, the expected angular size of the jet is thus $\lesssim 2$~arcsec.

For comparison, the observed spectrum of the radio emission from \cfhqs\ between $\nu_{\rm min}=120$--168~MHz and $\nu_{\rm max}=32$~GHz can be approximately described by a power law with $\alphar= 0.7\pm 0.3$ ($dF_\nu/d\nu\propto \nu^{-\alphar}$, see \citealt{Medvedev2020,Shao2020} and references therein), i.e. it is significantly flatter than the X-ray spectrum. The flux density at the lowest available frequency $\nu_{\rm min}$ is $S_\nu\approx 10$~mJy \citep{Shimwell2019}. This implies that the synchrotron spectrum peaks at $\nu_{\rm p}<120$~MHz and has a maximum intensity $S_{\rm p}>10$~mJy. Based on the standard synchrotron self-absorption argument, the required magnetic field $B\sim f(\gamma)^{-5}\theta^4\nu_{\rm p}^5 S_{\rm p}^{-2}(1+z)^{-1}$ \citep{Kellermann_1981}, where $f(\gamma)\sim 8$ is a weak function of $\gamma$, $\theta$ is the size of the radio source in milliarcseconds, and $\nu_{\rm p}$ and $S_{\rm p}$ are measured in GHz and Jy, respectively. The size of \cfhqs\ at the lowest radio frequencies is unknown, but the source has been resolved at 1.6~GHz with the inferred characteristic size $\approx 3$~mas \citep{Frey2011}.  This corresponds to a physical size of $\approx 20$~pc, which is a factor of $\lesssim 10^3$ shorter than the expected size of the X-ray jet (see above). We then find:
\begin{equation}
    B\sim 3\times 10^{-5}\left(\frac{\theta}{3~{\rm mas}}\right)^4\left(\frac{\nu_{\rm p}}{100~{\rm MHz}}\right)^5\left(\frac{10~{\rm mJy}}{S_{\rm p}}\right)^2~{\rm G}.
    \label{eq:bfield}
\end{equation}

This implies that the electrons responsible for the synchrotron radio continuum between 120~MHz and 32~GHz have a power-law distribution with a slope of $\pr=2.4\pm 0.6$ between $\gamma\sim 3\times 10^3$ and $\sim 5\times 10^4$ (for the fiducial parameter values in the equation above). Therefore, the electrons producing the radio emission are more energetic and have a flatter energy distribution than the electrons that presumably produce the X-ray emission via inverse Compton scattering of the CMB photons. 

The above consideration suggests the following possible scenario for \cfhqs. The quasar has been active for at least a few tens of thousands years, so that high-energy electrons have had enough time to cool down to $\gamma_{\rm min}$ (see the corresponding estimates above). A power-law distribution of electrons with $p\sim 2.4$ has been constantly supplied to the jets. The observed radio emission is produced via the synchrotron mechanism by freshly generated energetic electrons close to the central source, where there are also sufficiently strong magnetic fields. As the electrons propagate farther away from the central source, they are cooled via inverse Compton scattering of the CMB photons. The observed steep X-ray spectrum reflects the modified population of electrons, which results from the balance between injection and inverse-Compton energy losses of electrons. The theoretically expected difference between the modified and initial slopes of the electron energy distribution $p^\prime-p=1$ \citep{Kardashev1962} is consistent, within the measurement uncertainties, with the value inferred here for \cfhqs\ from the observed slopes of the X-ray and radio spectra: $\px-\pr=1.6\pm 0.7$.

Is this scenario self-consistent? First of all, we should check if inverse Compton scattering of emission from the accretion disc of the central supermassive black hole could cause additional significant energy losses for the electrons in the jets. Assuming that the central source is point-like, we may estimate the radiation energy density at distance $r$ from it in the comoving frame of the relativistic jet as $U_{\rm QSO}\sim L_{\rm QSO}/16\pi \Gamma^2cr^2$ (e.g. \citealt{dermer1993}). This should be compared with the energy density of the CMB as seen by the jet, $U_{\rm CMB}=4\sigma T_{\rm CMB}^4(1+z)^4\Gamma^2/c$. Taking the bolometric luminosity of the central source to be $L_{\rm QSO}\sim 2\times 10^{47}$~erg~s$^{-1}$ (see above), we infer that the CMB starts to dominate at $\sim 140$~pc from the black hole. 
 The jet propagation time out to this distance is $\sim 500$~yr, which is shorter by a factor of $\sim 20$ than the characteristic cooling time (at this distance) $t_{\rm cool} (\gamma_{\rm max})\sim 10^4$~yr of the highest energy electrons presumably responsible for the observed X-ray radiation (see eq.~[\ref{eq:tcool}]). This suggests that the quasar's optical--X-ray radiation probably plays an important role in cooling the electrons in the inner regions of the jets (and leads to the production of gamma-ray emission as a result), but it should not dramatically affect our conclusion that kpc-scale X-ray jets can form in this system.

We should next compare the synchrotron and inverse Compton energy losses. From equation~(\ref{eq:bfield}), for the quoted fiducial values of the parameters we find that the magnetic energy density within the (radio) jets $U_B=B^2/8\pi\sim 4\times 10^{-11}$~erg~cm$^{-3}$ is a factor of $\sim 10^3$ smaller than the CMB energy density in the comoving frame ($U_{\rm CMB}\sim 3\times 10^{-8}$~erg~cm$^{-3}$). This is largely consistent with the observed proportion of the radio and X-ray luminosities of \cfhqs=\srge: $\sim {\rm a~few~} 10^{43}$ and $\sim {\rm a~few~} 10^{46}$~erg~s$^{-1}$, respectively \citep{Medvedev2020}. 

We can further estimate the energy density of the radio photons within the jet, based on the luminosity and angular size of the radio source as $U_{\rm radio}\sim {\rm a~few~} 10^{-9}$~erg~s$^{-1}$, which is an order of magnitude smaller than $U_{\rm CMB}$. Therefore, the synchrotron self-Compton mechanism is probably a minor contributor to the observed X-ray emission.

We thus conclude that the inverse Compton/CMB/jet scenario is a feasible one for explaining the unusual properties of \cfhqs. The object may prove to be the most spectacular example so far of an anticipated population (e.g. \citealt{ghisellini_2014}) of high-redshift quasars with bright X-ray/weak radio jets. So far only a few such objects have been found via direct X-ray/radio imaging \citep{simionescu2016,schwartz2020}. In these objects, at redshifts between 2.5 and 4.7, the inferred jet/quasar luminosity ratio ranges from $\sim 1$ to $\sim 20$ per cent. In the case of \cfhqs, at $z=6.18$, we might be dealing with a more extreme case when the jet's luminosity (and hence its mechanical power) is comparable to that of the accretion disc. As discussed above, the expected size of the X-ray jet in this scenario is $\lesssim 2$~arcsec. Therefore, there is a hope that it could be spatially resolved with the {\it Chandra X-ray Observatory}, which would be a crucial observation for understanding the nature of high-redshift radio-loud quasars.

\section*{Acknowledgements}
\addcontentsline{toc}{section}{Acknowledgements}
This work is based on observations obtained with {\it XMM-Newton}, an ESA science mission with instruments and contributions directly funded by ESA Member States and NASA. This work was partially supported by the Russian Science Foundation, project \#19-12-00369. 
We made use the corner Python module\footnote{\url{https://github.com/dfm/corner.py}} for visualizing the posterior probability distributions \citep{corner}.  We are grateful to the anonymous referee  for their constructive  comments and suggestions.


\section*{Data Availability}

The data used here is publicly available in ESA’s XMM-Newton data archive. 



\bibliographystyle{mnras}
\bibliography{xmm_cfhqs} 



\bsp	
\label{lastpage}
\end{document}